\documentclass[pra,twocolumn,preprintnumbers,amsmath,amssymb,nofootinbib,showpacs]{revtex4}
\usepackage{graphicx}
\usepackage{amssymb}
\usepackage{mathrsfs}
\renewcommand\d{\partial}

\newcommand{\beq}{\begin{equation}}
\newcommand{\eeq}{\end{equation}}

\begin{document}
\preprint{TAUP-2965/13, NT@UW-13-18, EFI-13-9}

\title{Effective theory of chiral two-dimensional superfluids}

\author{Carlos Hoyos}
\affiliation{Raymond and Beverly Sackler Faculty of Exact Sciences,
School of Physics and Astronomy,
Tel-Aviv University, Ramat-Aviv 69978, Israel}
\author{Sergej Moroz}
\affiliation{Department of Physics, University of Washington,
Seattle, Washington 98195, USA}
\author{Dam Thanh Son}
\affiliation{Enrico Fermi Institute, James Franck Institute, and Department
of Physics, University of Chicago, Chicago, Illinois 60637, USA}

\date{May 2013}

\begin{abstract}
We construct, to leading orders in the momentum expansion, an
effective theory of a chiral ($p_x+ip_y$) two-dimensional fermionic
superfluid at zero temperature that is consistent with
nonrelativistic general coordinate invariance. This theory naturally incorporates the parity and time-reversal violating effects such as the Hall viscosity and the edge current. The  particle number current and stress
tensor are computed and their linear response to electromagnetic
and gravitational sources is calculated.  We also consider an isolated
vortex in a chiral superfluid and identify the leading chirality
effect in the density depletion profile.
\end{abstract}


\maketitle

\section{Introduction}
Chiral fermionic superfluids enjoy a long-standing and continuous
interest in condensed matter physics. This goes back to studies of
thin films of $^{3}\text{He-A}$ which is believed to form a chiral
condensate \cite{Vollhardt1990, Volovik2009}. More recently, low-dimensional chiral
superfluids attracted some attention due to the presence of Majorana
zero-energy edge quasiparticles \cite{Qi2011,Alicea2012} that might
facilitate progress towards a fault-tolerant quantum computer
\cite{Kitaev2003,Nayak2008}.

In this paper we construct and analyze the effective theory of a
two-dimensional fermionic superfluid at zero temperature that forms a
chiral condensate which in momentum space takes the form
\beq \label{porder}
\Delta_{\mathbf{p}}=(p_1\pm i p_2)\hat\Delta,
\eeq
where $\hat\Delta$ is a real function of $|\mathbf{p}|$ and the
sign defines the chirality of the condensate. In addition, $p_1$ and $p_2$ are projections of the momentum vector on the orthonormal spatial vielbein which will be defined later. The order parameter is
the p-wave eigenstate of the orbital angular momentum.  
We will assume that~(\ref{porder}) is the energetically favored 
ordering.
The usual spontaneous breaking of
the global particle number $U(1)_{N}$ symmetry is accompanied by the breaking
of the vielbein rotation $SO(2)_V$ symmetry. The
condensate \eqref{porder} remains invariant under the diagonal
combination of $U(1)_N$ and $SO(2)_V$ transformations which leads to
the symmetry breaking pattern
\beq \label{SSB}
U(1)_N\times SO(2)_V\rightarrow U(1)_D.
\eeq 
This implies the presence of a single gapless Goldstone mode in the
spectrum that governs the low-energy and long-wavelength dynamics of
the superfluid at zero temperature. The effective theory of this
Goldstone field has an infinite number of terms and can be organized
in a derivative expansion. This allows to include corrections to the well-known Landau superfluid hydrodynamics in a systematic fashion and to study phenomena at length scales that are larger than microscopic length scales (e.g. the coherence length). In this paper we restrict our attention
only to the leading and next-to-leading order terms in the derivative expansion.

Our guiding principle is the nonrelativistic version of general coordinate invariance developed in \cite{Son2006b}. We put the chiral superfluid into a curved space, switch on an electromagnetic source and demand the invariance of the effective theory with respect to nonrelativistic diffeomorphisms and $U(1)_N$ gauge transformations. Since the general coordinate invariance can be viewed as a local version of Galilean symmetry, the symmetry constraints on the effective theory are (even in flat space) more restrictive than just the ones imposed by the Galilean invariance alone.  This approach proved to be useful before and led to new predictions for unitary fermions \cite{Son2006b,Son2007} and quantum Hall physics \cite{Hoyos2012}.

While our effective theory might be viewed as a toy model mimicking
only some aspects of low-energy dynamics in thin films of
$^{3}\text{He-A}$ and spin-triplet superconductors, our predictions,
in principle can be tested experimentally with ultracold
spin-polarized fermions confined to two spatial dimensions.

\section{Effective theory of chiral superfluids}
Our starting point is the effective theory of a conventional (s-wave) superfluid constructed in \cite{Son2006b}. In curved space with a metric $g_{ij}$ and in the presence of an electromagnetic $U(1)_N$ source $A_{\nu}$, the leading order action of the Goldstone field $\theta$ was found to be\footnote{In this paper we follow the notation of \cite{Son2006b} except that the mass of the fermion is set to unity.}
\beq \label{LOS}
S[\theta]=\int dt d \mathbf{x} \sqrt{g} P\left( X\right),
\eeq
where $g\equiv \text{det} g_{ij}$,
\beq \label{X}
X=\mathscr{D}_t \theta -\frac{g^{ij}}{2} \mathscr{D}_i \theta\mathscr{D}_j \theta
\eeq
and the covariant derivative $\mathscr{D}_{\nu} \theta\equiv \partial_{\nu}\theta-A_{\nu}$ with $\nu=t, x, y$. The superfluid ground state has a finite density, that we characterize by the chemical potential $\mu$. In the effective action it enters as a background value for the Goldstone field, that is decomposed as
\beq \label{theta}
\theta=\mu t-\varphi
\eeq
with $\varphi$ standing for a phonon fluctuation around the ground state. This allows to identify the function $P(X)$ in Eq. \eqref{LOS} with the thermodynamic pressure as a function of the chemical potential $\mu$ at zero temperature. As demonstrated in \cite{Son2006b}, the expression $X$ transforms as a scalar under the infinitesimal nonrelativistic diffeomorphism transformation $x^i\to x^i+\xi^i$ provided
\beq
\begin{split}
\delta \theta&=-\xi^k \partial_k \theta, \\
\delta A_t&=-\xi^k \partial_k A_t-A_k \dot{\xi}^k, \\
\delta A_i&=-\xi^k \partial_k A_i-A_k \partial_i \xi^k+g_{ik}\dot{\xi}^k, \\
\delta g_{ij}&=-\xi^k \partial_k g_{ij}-g_{ik}\partial_j \xi^k-g_{kj}\partial_i \xi^k.
\end{split}
\eeq
In addition, $X$ is invariant under $U(1)_N$ gauge transformations.
These observations ensure that the leading order effective action \eqref{LOS} is invariant under general coordinate transformations.

As becomes clear from the symmetry breaking pattern \eqref{SSB}, in the case of a chiral superfluid we need a gauge potential for $SO(2)_V$ vielbein rotations. To this end we introduce an orthonormal spatial vielbein $e^{a}_{i}$ with $a=1, 2$. It is easy to check that the vielbein necessarily satisfies\footnote{Here we introduced the antisymmetric Levi-Civita symbol $\epsilon^{ij}=\epsilon_{ij}$, $\epsilon_{12}\equiv+1$. The Levi-Civita tensor is then $\varepsilon^{ij}=\frac{1}{\sqrt{g}}\epsilon^{ij}$, $\varepsilon_{ij}=\sqrt{g}\epsilon_{ij}$.} 
\beq \label{vielcons}
 g_{ij}=e^{a}_{i} e^{a}_{j}, \quad \epsilon^{ab}e^{a}_{i}e^{b}_{j}=\varepsilon_{ij}.
 \eeq
Such vielbein however is not unique and defined only up to a local $SO(2)_V$ rotation in the vielbein index $a$. This allows us to introduce a connection
\beq \label{NRomega}
\begin{split} 
\omega_t&\equiv\frac{1}{2}\Big(\epsilon^{ab}e^{aj} \partial_t e^{b}_{j}+B \Big), \\
\omega_i&\equiv\frac{1}{2}\epsilon^{ab}e^{aj} \nabla_i e^{b}_{j}=\frac{1}{2}\Big(\epsilon^{ab}e^{aj} \partial_i e^{b}_{j}-\varepsilon^{jk}\partial_j g_{ik} \Big),
\end{split}
\eeq
where $e^{aj}\equiv e^{a}_{i}g^{ij}$ and the magnetic field $B\equiv\varepsilon^{ij}\partial_i A_j$. Under a local (i.e., time- and position-dependent) infinitesimal $SO(2)_V$ rotation
\beq
e^{a}_{i}\to e^{a}_{i}+\phi(t,\mathbf{x})\epsilon^{ab}e^{b}_{i},
\eeq
the connection $\omega_{\nu}$ transforms as an Abelian gauge field $\omega_{\nu}\to \omega_{\nu}-\partial_{\nu} \phi$. Using Eq. \eqref{NRomega} together with the transformation law of the vielbein one-form
\beq
\delta e^{a}_i=-\xi^k \partial_k e^{a}_i-e^a_k \partial_i \xi^k, \\
\eeq
one can show that $\omega_{\nu}$ transforms as a one-form under the nonrelativistic diffeomorphisms, i.e.,
\beq
\delta \omega_{\nu} =-\xi^k \partial_k \omega_{\nu} -\omega_k \partial_{\nu}\xi^k.
\eeq 
In hindsight, this simple transformation of $\omega_{\nu}$ clarifies the appearance of the magnetic field $B$ in the definition of $\omega_t$.

We are now in position to construct the leading order effective theory of a chiral superfluid. Quite naturally, the theory is still defined by the action \eqref{LOS} with the covariant derivative now given by
\beq \label{covder}
\mathscr{D}_{\nu} \theta\equiv \partial_{\nu}\theta-A_{\nu}-s \omega_{\nu}.
\eeq
For the chiral p-wave superfluid $s=\pm 1/2$ with the sign determined by chirality of the ground state.\footnote{The effective action previously found in \cite{Stone2004} is contained in Eq. \eqref{LOS}. This becomes obvious after expanding $P(X)$ around the ground state $X=\mu$. } This ensures that the Goldstone boson is coupled to the proper (broken) linear combination of $U(1)_N$ and $SO(2)_V$ gauge fields. 
One must set $s=\pm n/2$ for a chiral superfluid with pairing in the $n^{\text{th}}$ partial wave.

It is interesting to note that instead of the usual formalism, where the superfluid is described by a
single Goldstone boson $\theta$, there is an alternative approach in
which the Lagrangian depends on the phase of the condensate $\theta$ and the superfluid velocity $v^i$. We present this alternative description of a (chiral) superfluid in Appendix \ref{alternative} and demonstrate there that by integrating
out $v^i$ one obtains the usual Goldstone effective action.

We point out that Galilean invariance\footnote{The infinitesimal Galilean boost with velocity $v^k$ is realized by a combination of the  gauge transformation $\alpha=v_k x^k$ together with the diffeomorphism $\xi^k=v^k t$.} alone is not sufficient to fix the leading order action \eqref{LOS}. Under a Galilean boost the magnetic field $B$ transforms as a scalar, i.e., $\delta B=-\xi^k \partial_k B$. For this reason the prefactor multiplying the magnetic field $B$ in the first equation in \eqref{NRomega} is not constrained by the Galilei invariance. Thus only the general coordinate invariance fixes uniquely the leading order Lagrangian of a chiral superfluid.

Time reversal and parity act nontrivially as
\beq
\begin{split}
T:& \, t\to -t, \, \theta\to -\theta, \, A_i\to -A_i, \, \omega_t\to -\omega_t; \\
P:& \,  x_1\leftrightarrow x_2, \, A_1\leftrightarrow A_2, \, \omega_t\to-\omega_t, \, \omega_1\leftrightarrow -\omega_2.
\end{split}
\eeq
For a fixed $s\ne 0$, the effective theory \eqref{LOS} is not separately invariant under neither time reversal $T$  nor parity $P$. On the other hand, $PT$ is a symmetry of the theory for any value of $s$. Note however that the action \eqref{LOS} is separately invariant under $T$ and $P$ if one transforms the chirality of the ground state, i.e.,  $s \to -s$. 

For $s=0$ the action \eqref{LOS} is the leading order term in a derivative
expansion that follows the counting in \cite{Son2006b}, where $\partial_{\nu} \theta\sim A_{\nu}\sim
g_{ij}\sim O(1)$ when expanded
around the ground state. In addition, in this power-counting a derivative
acting on any field $\mathscr{O}$ increases the order by one, i.e.,
$[\partial_{\nu}\mathscr{O}]=1+[\mathscr{O}]$. Non-linear effects of
fields with $[\mathscr{O}]=0$ are included, since
$[\mathscr{O}^n]=n[\mathscr{O}]=0$ for any $n\ge 1$. For this reason,
we conclude from the scaling of the metric that $[e^{a}_{i}]=0$ and
$[\omega_{\nu}]=1$. As a result, $\mathscr{D}_{\nu}\theta$ turns out
to be of a mixed order: it contains terms both of leading and
next-to-leading order in the derivative expansion.  Since one can show that the only possible next-to-leading order contribution consistent with symmetries is already contained in $\mathscr{D}_{\nu}\theta$, our theory is complete up to and including next-to-leading order.\footnote{The term in the Lagrangian of the form $Q(X)(\partial_t+v^i \partial_i)X$ is consistent with continuous symmetries for an arbitrary function $Q(X)$ and gives contributions to next-to-leading order. It does not however respect $PT$ invariance and should not be included.}

By introducing the superfluid density $\rho\equiv dP/dX$ and the superfluid velocity $v_j\equiv -\mathscr{D}_j \theta$ the nonlinear equation of motion for the Goldstone field can be written in the general covariant form
\beq \label{EOMgr}
\frac{1}{\sqrt{g}}\partial_t (\sqrt{g} \rho)+\nabla_i \left(\rho v^i \right)=0,
\eeq
which is the continuity equation in curved space.
With respect to nonrelativistic diffeomorphisms, $\rho$ transforms as a scalar and $v_i$ transforms as a gauge potential, i.e.,
\beq
\begin{split}
 \delta\rho&=-\xi^k \partial_k \rho, \\
 \delta v_i &=-\xi^k \partial_k v_i-v_k\partial_i \xi^k+ g_{ik}\dot{\xi}^k.
\end{split}
\eeq

By linearizing the equation of motion \eqref{EOMgr} in the absence of background gauge fields ($A_{\nu}=\omega_{\nu}=0$) one finds the low-momentum dispersion relation of the Goldstone field to be
\beq
\omega^2=c_{s}^2 \mathbf{p}^2,
\eeq
where the speed of sound $c_s\equiv \sqrt{\partial P/\partial\rho}$ is evaluated in the ground state.

From Eq. \eqref{covder} the vorticity $\omega\equiv \frac{1}{2}\epsilon^{ij}\partial_i v_j$ can be expressed as
\beq
\omega=\frac{\sqrt{g}}{2}\left(B+\frac{s}{2}R \right),
\eeq
where  we used that the Ricci scalar $\frac{1}{2}\sqrt{g} R=\epsilon^{ij}\partial_i \omega_j$ in two dimensions.
In the absence of a magnetic field in flat space, the superfluid velocity field is irrotational, i.e., $\omega=0$, except for singular quantum vortex defects. For an elementary vortex located at a position $\mathbf{x}_{v}$ the vorticity is $\omega(\mathbf{x})=\pi \delta(\mathbf{x}-\mathbf{x}_v)/2$. 
Single-valuedness of the macroscopic condensate wave-function yields quantization of the total magneto-gravitational flux on a compact manifold. For example, for a p-wave superfluid living on a sphere $S^2$ with no magnetic field, the total flux is $\int_{S^2}\omega=\pi$, which is accommodated in two elementary quantum vortices.

\section{Current and stress tensor} \label{sec:currents}
In this section, starting from the effective action \eqref{LOS}, we construct and analyze the $U(1)_N$ current and the stress tensor.

\subsection{$U(1)_N$ current}
In curved space the temporal and spatial part of the $U(1)_N$ three-current are found to be
\beq \label{U1cur}
\begin{split}
J^t&\equiv-\frac{1}{\sqrt{g}}\frac{\delta S}{\delta A_t}=\rho \\
J^i&\equiv-\frac{1}{\sqrt{g}}\frac{\delta S}{\delta A_i}=\underbrace{\rho g^{ij}v_j}_{\text{convective}}+\underbrace{\frac{s}{2}\varepsilon^{ij}\partial_j \rho}_{\text{edge}}.
\end{split}
\eeq
In addition to the usual convective term, we find the parity-odd contribution to the current which is proportional and perpendicular to the gradient of the superfluid density. In a near-homogeneous finite system this part of the current flows along the edge of the sample, where the density changes rapidly. It is important to stress that the edge current is perpendicular to $\partial_j\rho$, but not to the electric field $E_{j}\equiv \partial_t A_j-\partial_j A_t$. Thus there is no static Hall conductivity in the chiral superfluid in agreement with general arguments of \cite{Read2000}. The edge current was found in the study of superfluid $^{3}\text{He-A}$ by Mermin and Muzikar \cite{Mermin1980}.

The current conservation equation is
\beq
\frac{1}{\sqrt{g}}\partial_t (\sqrt{g}J^t)+\nabla_i J^i=0.
\eeq
Since $\nabla_i J^{i}_{\text{edge}}=0$, this is consistent with the equation of motion \eqref{EOMgr}.

In a non-homogeneous chiral superfluid the edge current is flowing even in the ground state. This implies that the ground state has a nonvanishing angular momentum $L_{\text{GS}}$. In flat space this is given by
\beq \label{ang}
L_{\text{GS}}=\int d^2 x \epsilon_{kl}x^k J^l=s\underbrace{\int d^2x \rho}_{N}, 
\eeq
where $N$ denotes the total number of fermions. This result has a simple explanation in the limit of strong interatomic attraction, where the many-body fermionic system can be viewed as a Bose-Einstein condensate of tightly bound molecules.  In the chiral superfluid with pairing in the $n^{\text{th}}$ orbital wave every molecule has the intrinsic angular momentum $l=\pm n$ with the sign determined by the chirality of the ground state. In the BEC limit interactions between molecules are weak and thus the total angular momentum is just the sum of internal angular momenta of separate molecules. 
Note however that Eq. \eqref{ang} should be valid also away from the BEC limit as long as the density near the edge varies on length scale that is large compared to any microscopic length scale.

The result \eqref{ang} was derived for a finite droplet of the chiral superfluid, where the density decreases continuously to zero at the boundary. Let us consider now a superfluid confined in a solid vessel. In this case, when we vary the action, together with the bulk current \eqref{U1cur} we find an additional current that flows along the one-dimensional boundary. This current is localized at the boundary and is given by
\beq
J^{i}_{\text{boundary}}=\frac{s}{2}\rho \varepsilon^{ij}\hat n_j,
\eeq
where $\hat n$ is a unit outer-pointing normal vector. The boundary current flows in the opposite direction to the edge part of the bulk current \eqref{U1cur}. When the velocity is zero, the total current at the boundary is 
\beq
J^{i}_{\text{boundary}}+\int d^2 x\, J^i_{\text{edge}}=0.
\eeq
The value of the angular momentum is then the same as in Eq. \eqref{ang}. However, the density is not an analytic function at the boundary in the topologically non-trivial phase. Therefore, the total angular momentum of the superfluid in the ground state confined in a solid vessel may change. For a thorough investigation of this question we refer to \cite{Sauls}.

More general values of the parameter $s$ are also possible. Interesting examples are anyon superfluids of fractional statistical phase  \cite{Chen1989, Greiter1989}
\beq
\theta=\pi\left(1-\frac1n\right).
\eeq
When $n=1$ the anyon becomes a boson (scalar) and when $n\to\infty$ it becomes a fermion. Similarly to the chiral superfluid, for any $n>1$ time reversal and parity invariance are broken.
Expanding the action \eqref{LOS} to second order around the ground state we find a term
\beq
\Delta L=\frac{s}{2}\frac{\rho}{c_s^2}A_t B.
\eeq
This was identified as responsible for the Landau-Hall effect in the anyon superfluid  \cite{Chen1989,Greiter1989}. The coefficient of this term was determined in  \cite{Halperin1989} to be
\begin{equation}
\Delta L=\frac{e^2}{8\pi}\left(n-\frac{1}{n}\right)A_t B.
\end{equation}
For anyons, scale invariance fixes $c_s^2=\rho/2$ leading to the identification
\beq
s=\frac12\left(n- \frac{1}{n}\right).
\eeq
Therefore the orbital angular momentum is also fractional, as expected.


\subsection{Stress tensor}
 In curved space the invariance of the effective action with respect to an infinitesimal nonrelativistic diffeomorphism $\xi^i$ implies
\beq
S[\theta+\delta\theta, A_{\nu}+\delta A_{\nu}, g_{ij}+\delta g_{ij}, e_{i}^{a}+\delta e_{i}^{a}]=S[\theta, A_{\nu}, g_{ij}, e_{i}^{a}],
\eeq
and leads to the Euler equation
\beq \label{euler}
\frac{1}{\sqrt{g}}\partial_t (\sqrt{g}J_k)+\nabla_i T^{i}_{\,k}=E_k J^t+\varepsilon_{ik}J^i B.
\eeq
where $T^{i}_{\,k}\equiv T^{ij}g_{jk}$ and the contravariant stress tensor is defined by
\beq
T^{ij}\equiv \frac{2}{\sqrt{g}}\frac{\delta S}{\delta g_{ij}}.
\eeq
Generically the Euler equation is the conservation equation only if $E_k=0$ and $B=0$.

Now we wish to compute the stress tensor $T^{ij}$ for the chiral superfluid \eqref{LOS}. In this case,
the variation of the vielbein is related to the variation of the metric via
\beq \label{varviel}
 e^{a}_i \to e^{a}_{i}+\frac{1}{2} e^{aj} \delta g_{ij}+\delta\lambda \epsilon^{ab}e^{b}_{i}.
\eeq
In this way the first relation in Eq. \eqref{vielcons} is satisfied up to second order in $\delta g_{ij}$. There is an ambiguity in the transformation of the vielbein parametrized by $\delta\lambda$ which is related to the $SO(2)_V$ gauge freedom of the vielbein. In the following we set $\delta\lambda=0$.
For the contravariant components of the metric,
\beq 
g^{ij}g_{jk} =\delta^{i}_{k} \quad \rightarrow \quad \delta g^{ij}=-g^{il} g^{jm} \delta g_{lm},
\eeq
so one cannot use the metric to raise the indices of the metric variation.

First, consider the s-wave superfluid, i.e., set $s=0$. Using $\delta\sqrt{g}=\frac{1}{2}\sqrt{g}g^{ij}\delta g_{ij}$ we find
\beq
T_{s=0}^{ij}=\frac{2}{\sqrt{g}}\frac{\delta S}{\delta g_{ij}}=P g^{ij}+\rho v^i v^j
\eeq
which is the stress tensor of the ideal fluid.

For the chiral superfluid additional variation of the action arise from the variation of the connection \eqref{NRomega}.
Namely, we find
\beq
\begin{split}
\delta \omega_t=&-\frac{1}{4}\varepsilon^{in}g^{jk}\partial_t g_{nk}\delta g_{ij} -\frac{1}{4}B g^{ij}\delta g_{ij}, \\
 \delta \omega_l=&-\frac{1}{4}\varepsilon^{in}g^{jk}\partial_l g_{nk}\delta g_{ij}-\frac{1}{2}\varepsilon^{jk}\partial_j \delta g_{lk}\\
 &+\frac{1}{4}\varepsilon^{mk}\partial_m g_{lk} g^{ij}\delta g_{ij}.
\end{split}
\eeq
This leads to 
\beq
\delta S_{\text{ch}}\equiv \int dt d\mathbf{x} \sqrt{g}P' \frac{\partial X}{\partial \omega_{\mu}}\delta\omega_{\mu}
\eeq
which in detail is given by
\begin{widetext}
\beq \begin{split}
\delta S_{\text{ch}}=&\frac{s}{4}\int dt d\mathbf{x} \sqrt{g}\Big[P'\varepsilon^{in}g^{jk}\partial_t g_{nk}-P'\mathscr{D}^l\theta \varepsilon^{in}g^{jk}\partial_l g_{nk} \Big]\delta g_{ij} 
        +\frac{s}{2} \epsilon^{kj} \int dt d\mathbf{x} \partial_k (P' \mathscr{D}^i \theta) \delta g_{ij} \\
        &+\frac{s}{4}\int dt d\mathbf{x} \sqrt{g} P' g^{ij}\left[B+\varepsilon^{mk}\mathscr{D}^l\theta\partial_m g_{lk} \right]\delta g_{ij}.
\end{split}
\eeq
\end{widetext}
After a tedious but straightforward calculation we find
\beq \label{stress}
\begin{split}
\Delta T^{ij}_{\text{ch}}&\equiv \frac{2}{\sqrt{g}}\frac{\delta S_{\text{ch}}}{\delta g_{ij}}\\
         &=(v^i J^j_{\text{edge}}+v^j J^i_{\text{edge}})+T^{ij}_{\text{Hall}}-\frac{s^2}{4}\rho R g^{ij}
\end{split}         
\eeq
with
\beq \label{Hallvis}
T^{ij}_{\text{Hall}}=-\eta_H (\varepsilon^{ik}g^{jl}+\varepsilon^{jk}g^{il}) V_{kl}.
\eeq 
Here the strain rate tensor $V_{kl}\equiv \frac{1}{2}\left(\nabla_k v_l+\nabla_l v_k+\partial_t g_{kl} \right)$ and $\eta_H=-\frac{s}{2}\rho$. 
With respect to general coordinate transformations both $V_{kl}$ and $T^{ij}_{\text{Hall}}$ transform as tensors. $T_{\text{Hall}}^{ij}$ is known as the Hall viscosity part of the stress tensor and was discovered first in \cite{Avron1995,Avron1997}. It generically arises in two-dimensional many-body systems that break time reversal and parity \cite{Read2009,Read2011}. 

In flat space with the metric $g_{ij}=\delta_{ij}$ the force density arising from the Hall viscosity is given by 
\beq
f_{\text{Hall}}^i=-\partial_jT^{ij}_{\text{Hall}}=\eta_{H}\epsilon^{ij}\Delta v_j,
\eeq
and thus the net work per unit of time produced by the Hall viscosity in a region $\mathscr{S}$ surrounded by a boundary $\partial \mathscr{S}$ is
\beq
w=\eta_{H} \int_{\mathscr{S}} v_i \epsilon^{ij}\Delta v_j=\eta_{H}\oint_{\partial\mathscr{S}} n^k \epsilon^{ij}v_i \partial_k v_j,
\eeq
where $n^k$ is the normal vector to the boundary $\partial\mathscr{S}$.
We conclude that the Hall viscosity is dissipationless in the bulk of the region $\mathscr{S}$. Alternatively, its contribution to the bulk entropy production is vanishing since  $T^{ij}_{\text{Hall}} V_{ij}=0$.

\section{Linear response}
In linear response theory the induced current $\delta \mathscr{J}$ is linear in the source $\delta\mathscr{A}$, i.e., 
\beq
\delta \mathscr{J}^{\mu}(t,\mathbf{x})=\int d \mathbf{x}' \int_{t'<t} dt' \mathscr{K}^{\mu\nu}(t,\mathbf{x};t',\mathbf{x}')\delta \mathscr{A}_{\nu}(t',\mathbf{x}'),
\eeq
where we introduced the response kernel $\mathscr{K}^{\mu\nu}(t,\mathbf{x};t',\mathbf{x}')$ with $\mu,\nu=t,x,y$. In a spacetime homogeneous system $\mathscr{K}^{\mu\nu}(t,\mathbf{x};t',\mathbf{x}')=\mathscr{K}^{\mu\nu}(t-t',\mathbf{x}-\mathbf{x}')$ and it is convenient to transform to Fourier space, where $\delta \mathscr{J}^{\mu}(\omega,\mathbf{p})= \mathscr{K}^{\mu\nu}(\omega,\mathbf{p})\delta \mathscr{A}_{\nu}(\omega,\mathbf{p})$.
 
Following \cite{Ambegaokar1961,Arseev2006,Roy2008}, one can determine linear response of a superfluid in three steps:
\begin{itemize}
 \item First, solve the linearized equation of motion of the Goldstone field in the presence of the external source $\delta \mathscr{A}$.
 \item Second, substitute the solution $\theta(\mathscr{A})$ into the definition of the current $\mathscr{J}$.
 \item Third, determine the induced current $\delta \mathscr{J}$ as a function of the source $\delta\mathscr{A}$.
\end{itemize}
The response to electromagnetic sources depends on the electric and magnetic fields, $E_i$ and $B$. The response to gravitational sources has similar contributions, that depend on a parity odd ``electric'' field $E_{\omega i}=\partial_t\omega_i-\partial_i\omega_t$ and the scalar curvature $R$. In fact, the form of the covariant derivative \eqref{covder} implies that the full response depends on the combinations
\begin{equation}
E_i^{\rm tot}=E_i+s E_{\omega i}, \ \ B^{\rm tot}=B+\frac{s}{2}R.
\end{equation}
In this section we compute the linear response of the $U(1)_N$ current and stress tensor to the electromagnetic and gravitational source respectively. The calculation is done in a homogeneous chiral superfluid ground state in flat space with  no background $U(1)_{N}$ potential $A_{\nu}=0$.

\subsection{$U(1)_N$ current response to electromagnetic source}
In the presence of the electromagnetic source $A_{\nu}$, the linearized equation of motion for the phonon field $\varphi$, defined in Eq. \eqref{theta}, takes the form of the relativistic wave equation
\beq
\partial_{t}^2 \varphi- c_{s}^2 \Delta \varphi=-\partial_t \left(A_t+\frac{s}{2}B \right)+c_{s}^2 \partial_i A^i
\eeq 
which is solved in momentum space
\beq
\varphi(\omega, \mathbf{p})=\frac{-i\omega \left(A_t+\frac{s}{2}B \right)-ic_{s}^2 p_i A^i}{\omega^2-c_{s}^2 \mathbf{p}^2}.
\eeq

By substituting this solution into Eq. \eqref{U1cur} we find to linear order in the source
\beq \label{den}
\delta\rho|_{\theta(A)}\equiv\rho|_{\theta(A)}-\rho^{\text{GS}}=
 \rho^{\text{GS}}\frac{\Big(ip_i E^i+\frac{s}{2}\mathbf{p}^2B\Big)}{\omega^2-c_{s}^2\mathbf{p}^2},
 \eeq 
 \beq \label{cur}
\begin{split}
 J^{i}|_{\theta(A)}
 =&\sigma(\omega,\mathbf{p})E^i+i\left(1-\frac{s^2}{4}\frac{\mathbf{p}^2}{c_{s}^2} \right)\rho_{\text{L}}(\omega, \mathbf{p})\epsilon^{ij}p_j B \\
 &+\sigma_{H}(\omega,\mathbf{p})\epsilon^{ij}E_j
\end{split}
\eeq
with
\beq \label{response}
\begin{split}
  \sigma(\omega,\mathbf{p})&=\rho^{\text{GS}}\frac{i\omega}{\omega^2-c_{s}^2 \mathbf{p}^2}, \\
  \rho_{\text{L}}(\omega,\mathbf{p})&=\rho^{\text{GS}}\frac{-c_{s}^2}{\omega^2-c_{s}^2 \mathbf{p}^2}, \\
 \sigma_{H}(\omega,\mathbf{p})&=\frac{s \rho^{\text{GS}}}{2}\frac{-\mathbf{p}^2}{\omega^2-c_{s}^2 \mathbf{p}^2}.
\end{split}
\eeq 
Here the electric field is $E_j=-i(p_j A_t+\omega A_j)$, the magnetic field is $B=i\epsilon^{ij}p_i A_j$ and $\rho^{\text{GS}}$ denotes the superfluid density in the ground state.
One can check explicitly that the conservation law $-i\omega \delta \rho+i p_i J^i=0$ is satisfied. 

In essence, the first line of Eq. \eqref{cur} governs the electromagnetic response of a conventional two-dimensional (s-wave) superfluid with the idealized Drude dynamical conductivity $\sigma(\omega, \mathbf{p})$ and the London diamagnetic response function $ \rho_{\text{L}}(\omega,\mathbf{p})$. Note that the numerical prefactor of the parity-even term $\sim  s^2$ in the bracket in front of the London term is not a reliable prediction of the theory \eqref{LOS} since it might be modified by additional next-to-next-to-leading order terms in the Lagrangian not discussed here. On the other hand, the second line of Eq. \eqref{cur} is responsible for the (anomalous) dynamical Hall response of the chiral superfluid. The peculiar property of the Hall conductivity $\sigma_{H}(\omega,\mathbf{p})$ is that it vanishes in the static limit, defined by setting first $\mathbf{p}=0$ and then taking $\omega\to 0$. In this strict sense the Hall conductivity of a chiral superfluid vanishes, which is in agreement with the 
arguments presented in Sec \ref{sec:currents}.

\subsection{Stress tensor response to gravitational source}
Here we consider a chiral superfluid in curved space with spatial metric $g_{ij}=\delta_{ij}+h_{ij}$ where  $h_{ij}$ is  a small perturbation around the flat background. The linearized equation of motion for the phonon field is given by
\beq
\partial_{t}^2 \varphi- c_{s}^2 \Delta \varphi=-s\partial_t \omega_t+c_{s}^2\partial_t h/2+s c_{s}^2\partial_i \omega^i,
\eeq
where $h\equiv \text{Tr} h_{ij}=h_{11}+h_{22}$. This can be solved in the Fourier space with the result
\beq
\varphi=\frac{i c_{s}^2 \omega h/2-is \omega \omega_t-is c_{s}^2 p_i \omega^i}{\omega^2-c_{s}^2 \mathbf{p}^2}.
\eeq
First, we find
\beq \label{help1}
\begin{split}
v_i&=i p_i\varphi+s\omega_i\\
     &=s\frac{\sigma(\omega, \mathbf{p})}{\rho^{\text{GS}}}E_{\omega i}+ic_{s}^2 \frac{\sigma(\omega, \mathbf{p})}{\rho^{\text{GS}}} p_i \frac{h}{2}+i\frac{s}{2} \frac{\rho_{\text{L}}(\omega, \mathbf{p})}{\rho^{\text{GS}}}\epsilon_{ij}p^j R,
\end{split}
\eeq
\beq \label{help2}
\delta P|_{\theta(g)}\equiv P|_{\theta(g)}-P^{\text{GS}}=c_{s}^2 \delta \rho|_{\theta(g)},
\eeq
where the parity-odd ``electric'' field $E_{\omega i}\equiv-i(\omega \omega_i+p_i \omega_t)$. As explained in Appendix \ref{interpret}, $E_{\omega i}$ is a gradient of the vorticity of the displacement field.   The variation of the pressure in Eq. \eqref{help2} follows from the thermodynamic definition of the speed of sound $c_{s}^2=\delta P/\delta \rho$. Now we substitute Eqs. \eqref{help1}, \eqref{help2} into the stress tensor \eqref{stress}. To linear order in sources the variation of the pure contravariant stress tensor can be written in a matrix form
\begin{widetext}
\beq \label{stresslin}
\begin{split}
\delta T|_{\theta(g)}\equiv&T|_{\theta(g)}-T^{\text{GS}}\\=&\big(\delta P|_{\theta(g)}-\frac{s^2}{4}\rho^{\text{GS}}R \big) \sigma_0+P^{\text{GS}}\delta g^{-1}+\\
&\eta_{H}\Big[-i\omega \frac{h_{xx}-h_{yy}}{2}-\frac{c_{s}^2}{2}\frac{\sigma(\omega, \mathbf{p})}{\rho^{\text{GS}}}(p_{x}^2-p_{y}^2) h +is\frac{\sigma(\omega, \mathbf{p})}{\rho^{\text{GS}}}(p_x E_{\omega x}-p_y E_{\omega y})-s\frac{\rho_{\text{L}}(\omega, \mathbf{p})}{\rho^{\text{GS}}}p_{y}p_{x}R \Big]\sigma_1+ \\
&\eta_{H}\Big[i\omega h_{xy}+c_{s}^2\frac{\sigma(\omega, \mathbf{p})}{\rho^{\text{GS}}}  p_x p_y h-is\frac{\sigma(\omega, \mathbf{p})}{\rho^{\text{GS}}}(p_x E_{\omega y}+p_y E_{\omega x})+\frac{s}{2}\frac{\rho_{\text{L}}(\omega, \mathbf{p})}{\rho^{\text{GS}}}(p_{y}^{2}-p_{x}^{2})R  \Big]\sigma_3,
\end{split}
\eeq 
\end{widetext}
where $\sigma_0$ is the unity matrix and $\sigma_{i}$ are the Pauli matrices. The first two terms in the square brackets of Eq. \eqref{stresslin} are linear in $s$ and break parity, we will discuss them in more detail below. The last two terms are proportional to $s^2$ and do not break parity. They are of higher order and might get corrections from next-to-next-to-leading order terms in the Lagrangian and thus are not predicted reliably by our theory.

Now we are ready to extract the dynamic Hall viscosity from the linear response calculation. Indeed, as argued in \cite{Saremi2012}, the gravitational wave
\beq
h_{ij}(t)=h_{ij}\exp(-i\omega t)
\eeq
induces the following perturbation of the off-diagonal component of the contravariant stress tensor
\beq
\delta T^{xy}=-P^{\text{GS}} h_{xy}+i\omega \eta(\omega)h_{xy}-i\omega \frac{\eta_H(\omega)}{2}(h_{xx}-h_{yy}),
\eeq 
where $\eta(\omega)\equiv \eta(\omega, \mathbf{p}=0)$ and $\eta_{H}(\omega)\equiv \eta_{H}(\omega, \mathbf{p}=0)$.
Direct comparison with Eq. \eqref{stresslin} gives us
\beq
\eta(\omega)=0,
\eeq
i.e., there is no shear viscosity because at zero temperature superfluid does not dissipate energy. In addition we find
\beq \label{visdyn}
\eta_{H}(\omega)=\eta_{H}=-\frac{s}{2}\rho^{\text{GS}}.
\eeq
It is natural that in the leading-order theory \eqref{LOS} the Hall viscosity does not depend on frequency.

\subsection{Relation between Hall conductivity and viscosity}
It was demonstrated in \cite{Hoyos2012} that for Galilean-invariant quantum Hall states the static electromagnetic Hall response at small momenta $\mathbf{p}$ receives a contribution from the Hall viscosity. Subsequently, the relation between the Hall conductivity and stress response was generalized to other Galilean-invariant parity-violating systems \cite{Bradlyn2012}. In addition, it was shown in \cite{Bradlyn2012} that the relation can be extended to all frequencies. In particular, for a chiral superfluid one finds
\beq
\eta_{H}(\omega)=\frac{\omega^2}{2}\frac{\partial^2}{\partial p_{x}^2}\sigma_{H}(\omega,\mathbf{p})\Big|_{\mathbf{p}=0}.
\eeq 
Using Eqs. \eqref{response} and \eqref{visdyn}, we checked that this relation is satisfied within our leading-order effective theory.

\subsection{Parity breaking terms and Kubo formula}
The parity breaking contributions to the stress tensor due to gravitational sources can be grouped in the odd viscosity tensor, defined through the linear response formula
\begin{equation}
T^{ij}_{\rm odd}=-\eta_{\rm odd}^{ijkl}\partial_t h_{kl}.
\end{equation}

The odd viscosity can be obtained from the Kubo formula involving the parity-odd part of the retarded two-point function of the stress tensor \cite{Bradlyn2012}. We 
define the frequency and momentum-dependent odd viscosity
\begin{equation}
\eta_{\rm odd}^{ijkl}(\omega,\mathbf{p})\equiv \frac{1}{i\omega^+}G_R^{ijkl}(\omega^+,\mathbf{p})_{\rm odd},
\end{equation}
where $\omega^+\equiv\omega+i\epsilon$ with $\epsilon\to 0$. In fact, the total viscosity tensor contains an additional contact term inversely proportional to the compressibility \cite{Bradlyn2012}, but it is parity-even and thus does not affect our discussion of the odd viscosity.

The retarded two-point function is
\begin{equation}
\begin{split}
&G_R^{ijkl}(\omega,\mathbf{p})=\\
&i\int_0^\infty dt \int d^2 x\,e^{i\omega t-i \mathbf{p}\cdot\mathbf{x}}\left\langle \left[\tau^{ij}(t,\mathbf{x}), \tau^{kl}(0,\mathbf{0})\right]\right\rangle,
\end{split}
\end{equation}
where $\tau^{ij}$ is the stress tensor in the absence of gravitational sources. 

To leading order in the phonon fluctuation, the parity even and odd contributions to the stress tensor are
\begin{equation}
\begin{split}
\tau^{ij}_{\rm even} &=P^{\text{GS}} \delta^{ij}-\rho^{\text{GS}} \delta^{ij}\partial_t\varphi,\\
\tau^{ij}_{\rm odd}  &=-\eta_H(\epsilon^{ik}\delta^{jl} +\epsilon^{jk}\delta^{il})\partial_k\partial_l\varphi.
\end{split}
\end{equation}
The parity-odd contributions to the two-point function come from cross terms of even and odd contributions. Therefore, to leading order
\begin{equation}
\begin{split}
&\left\langle \tau_{\rm odd} \tau_{\rm even}^{ij}\right\rangle =\rho^{\text{GS}} \eta_H\delta^{ij} \left[ \sigma^1(\partial_y^2-\partial_x^2)+2\sigma^3\partial_x\partial_y\right]\partial_t \langle\varphi\varphi\rangle,\\
&\left\langle \tau_{\rm even} \tau_{\rm odd}^{ij}\right\rangle =\rho^{\text{GS}} \eta_H\sigma^0 (\epsilon^{ik}\delta^{jl} +\epsilon^{jk}\delta^{il})\partial_k\partial_l\partial_t \langle\varphi\varphi\rangle,\\
\end{split}
\end{equation}
where we are using the same matrix notation as before for the first entry of the stress tensor in the two-point function. In addition, we used $\langle \varphi \rangle=0$.

Using the Fourier transform of the  two-point function of the phonon field
\begin{equation}
G^{\varphi\varphi}(\omega,\mathbf{p})=\frac{1}{\rho^{\text{GS}}}\frac{1}{\frac{1}{c_s^2}\omega^2-\mathbf{p}^2},
\end{equation}
one can recover the results we have derived before for the linear response of the stress tensor to gravitational sources.
Specifically, the odd-even contribution produce the terms proportional to the trace of the metric perturbation in the brackets in Eq. \eqref{stresslin}. These terms are especially interesting because they are related to the fact that the superfluid is compressible. The metric perturbation changes the volume form by a term proportional to its trace
$\delta\sqrt{g}=h/2$. This excites phonons that produce the stress we have computed above. The even-odd terms give contributions to the variation of the pressure.

The regular Hall viscosity term on the other hand does not involve the phonon propagator and therefore it is a kinematic response that will appear in the correlation function as a contact term. Its origin is similar to the diamagnetic current, which is obtained from a term in the action quadratic in sources.

\section{Vortex solution}
Here we consider a vortex in a chiral superfluid in flat space placed at the origin. We treat this problem in polar coordinates $(r,\phi)$. Far away from the core, due to the single valuedness of the condensate wave function
\beq \label{vel}
v_{r}=0, \qquad v_{\phi}=\frac{n}{2r}, \qquad n\in \mathbb{Z}.
\eeq 
Here $v_r$ and $v_{\phi}$ are  the coefficients in the decomposition $\mathbf{v}=v_r \mathbf{e}_r+ v_{\phi}\mathbf{e}_{\phi}$ with $\mathbf{e}_r$ and $\mathbf{e}_{\phi}$ denoting the unit vectors in the radial and angular directions.

We will determine the asymptotic behavior of the superfluid density $\rho(r)$ as $r\to\infty$.
From Eq. \eqref{euler} the static Euler equation reads
\beq
\rho v^j\partial_j v^i=-\partial^i p+\eta_H \epsilon^{ij}\Delta v_j-\partial_j ( v^i J^{j}_{\text{edge}}+v^j J^{i}_{\text{edge}})
\eeq
which after the projection onto the radial direction becomes
\beq
\frac{\rho}{r} v_{\phi}^2=c_{s}^2\partial_r \rho-\eta_{H}\big[\partial^2_r v_{\phi}+\frac{1}{r}\partial_r v_{\phi}-\frac{1}{r^2} v_{\phi} \big]-f_{\text{edge}},
\eeq
where $f_{\text{edge}}\equiv-\delta_{ik}e^{k}_r\partial_j ( v^i J^{j}_{\text{edge}}+v^j J^{i}_{\text{edge}})=-\frac{sn}{2r^2} \partial_r \rho$.  For the velocity given by Eq. \eqref{vel} we have $\mathbf{\nabla}\cdot \mathbf{v}=0$ and thus $\Delta \mathbf{v}=0$. This simplifies the previous equation to the form
\beq \label{denprof}
\rho\frac{n^2}{4r^3}=\Big[c_{s}^2+\frac{sn}{2r^2}\Big]\partial_r \rho.
\eeq

To leading order in the large-distance expansion
\beq
\Big[c_{s}^2+\frac{sn}{2r^2}\Big]\to c_{s\infty}^2\equiv \frac{\partial P}{\partial \rho}\Big|_{r\to\infty} 
\eeq
and the differential equation \eqref{denprof} simplifies to
\beq \label{leadingeq}
\frac{d\rho}{\rho}=\frac{n^2 dr}{4c_{s\infty}^2 r^3}.
\eeq 
It is easily integrated giving the density profile
\beq \label{leadingprof}
\rho(r)
=\rho_{\infty}\big[1-\frac{n^2}{8c_{s\infty}^2 r^2}+O(r^{-4}) \big],
\eeq
where $\rho_{\infty}$ is the asymptotic value of the superfluid density away from the vortex core.
Since the chirality parameter $s$ does not appear in Eq. \eqref{leadingeq}, the leading order tail of the density profile is invariant under $n\to -n$.

Chirality effects arise first at next-to-leading order in the large-distance expansion. By  using
\beq
c_{s}^2=\frac{\partial P}{\partial \rho}= c_{s\infty}^2+\frac{\partial^2 P}{\partial\rho^2}\Big|_{r\to\infty}(\rho-\rho_{\infty})+O\big[(\rho-\rho_{\infty})^2\big]
\eeq
together with Eq. \eqref{leadingprof}, we find that to next-to-leading order we can replace
\beq
\Big[c_{s}^2+\frac{sn}{2r^2}\Big]\to c_{s\infty}^2+\Big(sn-n^2\frac{\rho \frac{\partial^2 P}{\partial\rho^2}}{4 c_{s}^2}\Big|_{r\to\infty} \Big)\frac{1}{2r^2}
\eeq
in Eq. \eqref{denprof}. The solution of Eq. \eqref{denprof}  now gives the superfluid density profile up to next-to-leading order
\begin{widetext}
\beq \label{psol}
\rho(r)=\rho_{\infty}\Big[1-\frac{n^2}{8c_{s\infty}^2 r^2}+\frac{n^4\Big(1-\frac{\rho \frac{\partial^2 P}{\partial \rho^2}}{c_{s}^2}\Big|_{r\to\infty} \Big)}{128c_{s\infty}^4 r^4}+\frac{s n^3}{32 c_{s\infty}^4 r^4}+O(r^{-6}) \Big].
\eeq
\end{widetext}
Since in the chiral superfluid time reversal and parity are spontaneously broken, the density profile is not invariant under $n\to -n$. The leading parity-violating correction appears first in the $1/r^4$ tail. The relative difference between the densities of a vortex (with $n=1$) and an antivortex (with $n=-1$) is asymptotically given by
\beq \label{result}
\frac{\Delta\rho}{\rho_{\infty}}=\frac{s}{16 c_{s\infty}^4 r^4}+O(r^{-6}).
\eeq

From the point of view of the microscopic theory, our hydrodynamic description becomes reliable only for $r\gg \xi$, where $\xi$ is the coherence length of the chiral superfluid.
For weak interatomic attraction (BCS regime) the density depletion in the vortex core is small. The maximal depletion away from the core is
\beq \label{relden}
\frac{\Delta\rho}{\rho_{\infty}}\Big|_{\text{max}}^{BCS}\sim \frac{s}{16 c_{s\infty}^4 \xi^{4}}\ll 1
\eeq
since $c_{s\infty} \xi\gg 1$ in the BCS limit.
On the other hand,  for strong interatomic attraction the total density depletion in the core becomes large. In addition, near the Feshbach resonance we obtain
\beq
\frac{\Delta\rho}{\rho_{\infty}}\Big|_{\text{max}}^{res}\sim \frac{s}{16 c_{s\infty}^4 \xi^{4}}\sim 1.
\eeq
This happens because near the Feshbach resonance the only relevant scale is given by the atomic density\footnote{Due to the marginal nature of interactions in two spatial dimensions, the effective range can not be set to zero at the Feshbach resonance. The dependence of physical observables on this length scale is logarithmic, i.e. weak, and can thus be neglected.} and thus $c_{s\infty} \xi\sim 1$ in this regime \cite{Gurarie:2005,Botelho:2005,Cheng:2005}. 
In summary, our result \eqref{result} for the long-range depletion profile should be physically relevant in the strongly-interacting region near the Feshbach resonance which contains in particular the topological quantum phase transition and its neighborhood \cite{Read2000}. Note, however, that due to the divergence of the correlation length at the phase transition, our approach has limited applicability in the close vicinity of the phase transition point.

\section{Conclusion}
In this paper we constructed the leading-order low-energy and long-wavelength effective hydrodynamic theory of the simplest chiral fermionic superfluid in two spatial dimensions. The effective theory incorporates time reversal and parity violating effects and thus naturally gives rise to the edge particle current and Hall viscosity. In agreement with \cite{Bradlyn2012}, we found a relation between the Hall conductivity and Hall viscosity response functions. As an application of the formalism, we constructed a quantum vortex solution and discovered that the leading chirality effect appears first in the $1/r^4$ tail of the density depletion.  Our predictions might be tested in experiments with spin-polarized two-dimensional ultracold fermions.

Since the chiral superfluid studied in this paper is a fermionic topological liquid, it is known to posses a protected gapless fermionic edge mode in the weakly-coupled regime \cite{Volovik2009, Read2000}.
Although our theory contains only the bosonic Goldstone mode as a degree of freedom, we believe that the fermionic gapless edge state is (at least partially) taken into account in our formalism. An important input for the effective theory is the exact equation of state $P(\mu)$ which defines the Lagrangian. Due to the topological phase transition, the function $P(\mu)$ is non-analytic at the critical value of the chemical potential $\mu_{cr}$. In mean-field studies one finds $\mu_{cr}=0$ \cite{Volovik2009, Read2000}. Consider now the ground state of a finite droplet of the chiral superfluid in the weakly-coupled regime. While in the bulk $\mu$ is almost constant and positive, it decreases towards zero near the boundary. This means that in the topologically non-trivial weakly-coupled phase there is a closed contour $\mu=\mu_{cr}$ which encircles the bulk, where the Lagrangian is non-analytic. This non-analyticity can appear only from integration of a massless mode. This one-dimensional closed contour is the 
domain where the gapless fermionic mode lives in the original fermionic model. On the other hand,  in the topologically trivial strongly-coupled regime there is no such a contour since in this regime $\mu<\mu_{cr}$ already in the bulk.  In summary, in the weakly-coupled regime the topologically protected fermionic gapless  mode was integrated out in Eq. \eqref{LOS} and manifest itself via the non-analyticity of the equation of state. In future it would be useful to derive the effective theory from the microscopic fermionic model.

It would be interesting to extend our theory to higher orders in the derivative expansion, where the nonrelativistic general coordinate invariance might put many more additional constraints compared to those imposed by considering Galilean invariance alone. Generalization of the theory to chiral superfluids with spin degrees of freedom might prove useful for better understanding of thin films of $^{3}\text{He}$.

\emph{Acknowledgments:}
The authors thank Tom\'a\v{s} Brauner,  Matthias Kaminski, Yusuke Nishida, Haruki Watanabe and Naoki Yamamoto for valuable discussions.  This work
was supported by  US DOE Grant
No.\ DE-FG02-97ER41014, and NSF MRSEC Grant No.\ DMR-0820054 and by the Israel Science Foundation (grant number 1468/06).
\appendix
\section{Alternative description of superfluid} \label{alternative}
We postulate that $v^\mu=(1, v^i)$ transforms like a vector under
spatial coordinate transformations, i.e., 
\beq
\delta v^\mu =
-\xi^\lambda\d_\lambda v^\mu + v^\lambda\d_\lambda \xi^\mu,
\eeq 
where $\xi^t=0$.  This means
\begin{equation}
\begin{split}
  \delta v^i &= -\xi^k \d_k v^i + v^k \d_k \xi^i + \dot \xi^i, \\
  \delta v_i &\equiv g_{ij} v^j=-\xi^k \partial_k v_i-v_k\partial_i \xi^k+ g_{ik}\dot{\xi}^k.
\end{split}   
\end{equation}
If such $v^i$ exists, then one can define the improved gauge potentials
\begin{align}
  \tilde A_t &\equiv A_0 + \frac{1}{2} g_{ij} v^i v^j, \\
  \tilde A_i &\equiv A_i - g_{ij} v^j,
\end{align}
so that $\tilde A_\mu$ transforms like a one-form 
\begin{equation}
  \delta\tilde A_\mu=-\xi^k\d_k \tilde A_\mu - \tilde{A}_k \d_\mu \xi^k.
\end{equation}

In this formalism the action of a conventional s-wave superfluid is
\begin{equation} \label{alt}
  S[\theta, v^i] =  \int\! dt d\mathbf{x}\,\sqrt g \left[\rho v^\mu (\d_\mu\theta - \tilde A_\mu)
      -\epsilon(\rho) \right],
\end{equation}
where $\epsilon(\rho)$ is the density of the internal energy that is not associated with the macroscopic motion of the superfluid.
After expanding $\tilde A_\mu$, this action becomes 
\begin{equation}
  S =  \int\! dt d\mathbf{x}\, \sqrt g\left[ \rho \mathscr{D}_t\theta  +\rho v^i \mathscr{D}_i\theta
      + \frac 12 \rho g_{ij} v^i v^j - \epsilon(\rho) \right]
\end{equation}
By integrating out $v^i$, we find that
\begin{equation}
  v^i = -g^{ij}\mathscr{D}_j\theta,
\end{equation}
i.e., $v^i$ is the superfluid velocity. Moreover, by integrating the theory \eqref{alt} over $v^i$ and
$\rho$ we reproduce the effective theory \eqref{LOS}.

To describe chiral superfluids in this formalism we need to use the improved connection
\begin{align}
  \omega_t &\equiv \frac12 \Big( \epsilon^{ab} e^{aj}\d_0 e^b_j +
      \varepsilon^{ij} \d_i v_j\Big),\\
  \omega_i &\equiv \frac12 \epsilon^{ab} e^{aj} \nabla_{\!i} e^b_j.
\end{align}
It can be checked that $\omega_\mu$ transforms like a one-form.  The
action of a chiral superfluid becomes
\begin{equation}
  S[\theta, v^i] =  \int\! dt d\mathbf{x}\,\sqrt g \left[\rho v^\mu (\d_\mu\theta - \tilde A_\mu
       - s\omega_\mu)
      - \epsilon(\rho) \right].
\end{equation}
Note that in this formalism the $U(1)_N$ current is pure convective
 \beq
 J^\mu\equiv-\frac{1}{\sqrt{g}}\frac{\delta S}{\delta A_\mu}=\rho v^\mu.
 \eeq

\section{Interpretation of $E_{\omega,i}$} \label{interpret}

In the linearized approximation around flat background $e_i^a=\frac{1}{2} h_i^a$, and one can check that the components of the connection \eqref{NRomega} are
\begin{equation}
\omega_t \sim O(h^2), \ \ \omega_i=-\frac{1}{2}\epsilon^{jk}\partial_j h_{ki}.
\end{equation}
The spatial metric can be interpreted as the stress source produced by a deformation $x^i\to x^i+\xi^i$, where $\xi^i$ is the displacement vector. In this case $h_{ij}=-\partial_{i}\xi_{j}-\partial_{j}\xi_{i}$, which is fixed by the transformation law of the metric under nonrelativistic diffeomorphisms. The spatial part of the connection \eqref{NRomega} now becomes
\begin{equation}
\omega_i=\frac{1}{2}\partial_i(\epsilon^{jk}\partial_j \xi_k).
\end{equation}
This can be viewed as the gradient of the torsion of the displacement field. To linear order, the parity-odd ``electric'' field is then
\begin{equation}
E_{\omega, i}\simeq \partial_t\omega_i=\partial_i\left(\frac{1}{2}\epsilon^{jk}\partial_j \partial_t \xi_k\right)\equiv \partial_i\Omega,
\end{equation}
where $\Omega$ is the vorticity of the displacement field.



\end{document}